\documentclass[aps,prl,preprint,superscriptaddress]{revtex4}

\usepackage[dvips]{graphicx}
\newcommand{\be}{\begin{displaymath}}
\newcommand{\bn}{\begin{equation}}

\newcommand{\en}{\end{equation}}
\newcommand{\ee}{\end{displaymath}}
\newcommand{\p}{\partial}

\begin{document}


\title{Stellarators with permanent magnets}



\author{P.~Helander}
\author{M.~Drevlak}
\affiliation{Max-Planck-Institut f\"ur Plasmaphysik, 17491 Greifswald, Germany}
\affiliation{Max Planck/Princeton Research Center for Plasma Physics}
\author{M.~Zarnstorff}
\author{S.C.~Cowley}
\affiliation{Princeton Plasma Physics Laboratory, New Jersey, USA}
\affiliation{Max Planck/Princeton Research Center for Plasma Physics}



\date{\today}

\begin{abstract}

It is shown that the magnetic-field coils of a stellarator can, at least in principle, be substantially simplified by the use of permanent magnets. Such magnets cannot create toroidal magnetic flux but they can be used to shape the plasma and thus to create poloidal flux and rotational transform, thereby easing the requirements on the magnetic-field coils. As an example, a quasiaxisymmetric stellarator configuration is constructed with only 8 circular coils (all identical) and permanent magnets.

\end{abstract}


\maketitle

\normalsize

Stellarators, tokamaks and other devices for fusion plasma confinement use electromagnets to create the magnetic field. In the case of stellarators, the required magnetic-field coils can be very complicated and contribute significantly to the overall cost of the device \cite{Bosch}. In the present Letter, we suggest that permanent magnets could be used to shape the plasma and drastically simplify the coils. Our emphasis is on mathematical aspects of this problem whereas other issues will be discussed in subsequent papers. These issues include properties of permanent magnets and why certain types are particularly suitable for stellarators, questions of engineering, assembly, and practical limitations. 

A magnetic field ${\bf B}$ tracing out toroidal surfaces can never be created by permanent magnets alone, because it follows from Amp{\`e}re's law that the line integral of the magnetic field taken once toroidally around the torus is proportional to the linked current of {\em free} charges,
	$$ \oint_C {\bf B} \cdot d{\bf r} = \mu_0 I_{\rm free}. $$
This conclusion follows from one of Maxwell's equations, 
	$$ \nabla \times {\bf B} = \mu_0 ({\bf J}_{\rm free} + \nabla \times {\bf M} ), $$
if the integration contour $C$ is chosen to lie within the plasma, where the magnetization ${\bf M}$ vanishes. In other words, permanent magnets cannot create a net toroidal magnetic flux, but they {\em can} (perhaps surprisingly) create poloidal flux and thus twist the magnetic field lines in a stellarator (though not in an axisymmetric device such as tokamak).  

To see how this can be accomplished, we write the magnetic field as a sum $ {\bf B} = {\bf B}_{\rm c} + {\bf B}_{\rm m}$, 
where 
	$$ {\bf B}_{\rm c}({\bf r}) = \frac{\mu_0}{4 \pi} \int_{\rm coils} {\bf J}_{\rm free}({\bf r}')
	\times \frac{{\bf r} - {\bf r}'}{|{\bf r} - {\bf r}'|^3} dV' $$
represents the field created by coils and ${\bf B}_{\rm m}$ that from the permanent magnets. The magnetization ${\bf M}$ vanishes outside a bounded domain $\Omega$ but is generally finite on the boundary $\p \Omega$ and produces a magnetic field
	\bn {\bf B}_{\rm m}({\bf r}) = \frac{\mu_0}{4 \pi} \left( \int_\Omega ( \nabla \times {\bf M} )
	\times \frac{{\bf r} - {\bf r}'}{|{\bf r} - {\bf r}'|^3} dV' \right. 
	 \left. +
	\int_{\p \Omega} ({\bf M} \times {\bf n})
	\times \frac{{\bf r} - {\bf r}'}{|{\bf r} - {\bf r}'|^3} dS' \right), 
	\label{BS}
	\en
where $\bf n$ is the unit vector pointing outward from $\Omega$. 

Our aim is to find a magnetization field $\bf M$ that creates a desired magnetic field ${\bf B}_m$ within the plasma region, which we denote by $P$. Since many different choices of $\bf M$ produce the same magnetic field, the solution is not unique and there is considerable freedom to find the simplest one. One way to solve the problem is to reduce it to one already routinely solved in stellarator design. This problem was first described by Merkel \cite{Merkel} and proceeds from the observation that the magnetic field in the plasma is uniquely determined by the shape of the plasma boundary $\p P$ and the current and pressure profiles within the plasma \cite{Kruskal-Kulsrud}. Suppose, therefore, that a desired plasma surface $\p P$ is prescribed and consider the problem of finding the surface current 
	\bn {\bf K} = {\bf n} \times \nabla \Phi 
	\label{K}
	\en
on another toroidal surface $\p D$, at some distance from the plasma, that creates a magnetic field tangential to $\p P$. In the method of Merkel, this is done by choosing the scalar function $\Phi$ on $\p D$ so as to minimize the surface integral
	\bn \int_{\p P} |{\bf n} \cdot {\bf B}|^2 dS.
	\label{Bn2}
	\en
(This problem is ill-posed but can be regularized in a number of ways, for instance by adding term proportional to $|{\bf K}|^2$ to the integrand \cite{Landreman}.) In conventional stellarator design, the surface current $\bf K$ thus found is subsequently discretized into suitable magnetic-field coils, but these are in general very complicated. 

To see how permanent magnets may help, it is useful to introduce a set of coordinates $(r,u,v)$ where $r=r_0$ is constant on $\p D$ and the other coordinates increase by unity in the poloidal and toroidal directions, respectively. 
The current potential $\Phi$ in Eq.~(\ref{K}) is in general of the form 
		$$ \Phi(u,v) = J u + Iv + \tilde \Phi(u,v), $$
where $\tilde \Phi(u,v)$ is a single-valued function on the surface $\p D$ in contrast to $\Phi$. ($\nabla \Phi$ is nevertheless single-valued.) $I$ and $J$ are constants proportional to the net currents in the toroidal and poloidal directions, which thus govern the topology of the coils. The constant $J$ vanishes for modular coils, and the constant $I$ determines the net toroidal magnetic flux inside $\p D$.  

If Merkel's problem is modified slightly by taking this net toroidal flux to be created by some given toroidal-field coils, then the net poloidal current on $\p D$ can be taken to vanish, $I = 0$. In this representation, the magnetic field is thus partly created by coils and partly by a surface current on $\p D$ having the property that $I=0$. The problem of finding a suitable magnetization field can then be reduced to Merkel's problem by choosing $\bf M$ in such a way that $\nabla \times {\bf M}$ vanishes within $\Omega$. The entire magnetization current entering in Eq.~(\ref{BS}) then appears as a surface current on $\p \Omega$, 
	$$ {\bf K} = {\bf M} \times {\bf n}. $$
As a result, ${\bf M}$ can be chosen as 
	\bn {\bf M} = -\nabla \left[ f(r,u,v) \tilde \Phi(u,v) \right], 
	\label{Merkel M}
	\en
where the function $f$ is equal to unity on the part of the boundary $\p \Omega_0$ that faces the plasma, which we identify with Merkel's current-carrying surface $\p D$. On the outward-facing boundary $\p \Omega_1$, we take $f$ to vanish.
For instance, if the domain $\Omega$ is the region $r_0 < r < r_1$, we take
	$$ f(r_0,u,v) = 1, $$
	$$ f(r_1,u,v) = 0. $$
With this prescription, the magnetization skin current, 
	\bn {\bf K} = {\bf n} \times (\tilde \Phi \nabla f + f \nabla \tilde \Phi) 
	= f {\bf n} \times \nabla \tilde \Phi, 
	\label{skin current}
	\en
becomes equal to that found by Merkel's procedure on $r=r_0$ and vanishes on $r=r_1$. 

If the width $r_1-r_0$ of the magnetization region is chosen to be large, the required magnetization $M = |{\bf M}|$ is relatively small but the volume occupied by magnets is large. If this volume is instead chosen to be small, the required magnetization is large. There is thus a basic trade-off between using a large volume of weak magnets or a small volume of strong ones. However, regardless of this choice of volume, the largest magnetization will always exceed the largest gradient of $\tilde \Phi$ on the plasma-facing surface,
	\bn M_{\rm max} \ge \max_{\p \Omega_0} | \nabla \tilde \Phi |. 
	\label{Mmax}
	\en
This condition places an important upper bound on the field strength achievable by arranging the magnets according to this method. 
	
\begin{figure}[htp]
    \centerline{\includegraphics[scale=0.4]{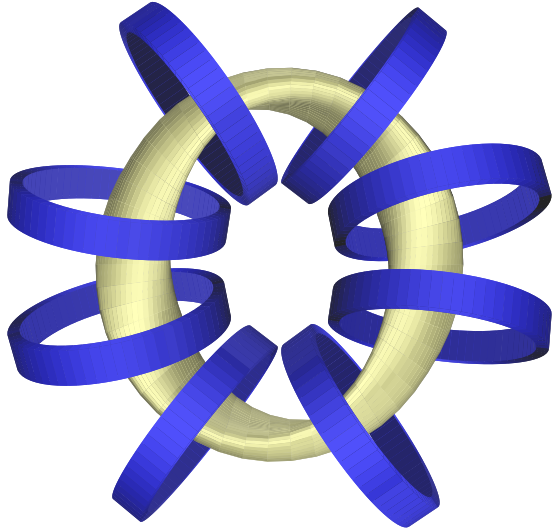}}
	\caption{\em Plasma boundary for ESTELL, a two-period quasisymmetric stellarator \cite{ESTELL}, shown with the proposed simplified set of 8 identical and circular coils, all carrying the same current.}
\end{figure}	
	
We now turn to a concrete example, an optimized stellarator configuration published a few years ago \cite{ESTELL}, which was orginally obtained by deforming a classical $l=2$ stellarator with aspect ratio $A=5$ into a shape that makes the magnetic field quasi-axisymmetric. (This means that the field strength is nearly independent of the toroidal angle in Boozer or Hamada coordinates, which ensures good orbit confinement \cite{JN,Boozer-PPCF,Helander-ROP}.) In the original design, the magnetic field was created by 20 non-planar, modular coils of 5 different types. Leaving permanent magnets to do most of the plasma shaping, a new optimization was now carried out where only 8 identical, planar, circular toroidal-field coils proved necessary. In this optimization, the orientation and positions of the coils were varied in such a way as to minimize Eq.~(\ref{Mmax}) under constraints ensuring that the coils do not get too close to each other or to the plasma. Accordingly, the resulting coils, which are shown in Fig.~1, are situated comfortably far from the plasma, ensuring a relatively small toroidal ripple. The magnetization surface current density in Eq.~(\ref{skin current}) required for plasma shaping is displayed in Fig.~2 and is well within the range achievable with Nd magnets. Here the device has been scaled to a field strength on the magnetic axis of $B_0 = 1$ T and an average major radius of $R=1.4$ m. The largest value of $\mu_0 \tilde \Phi$ is then about 0.25 Tm and the required thickness of the magnetization region is about 18 cm in the thickest regions, thus providing clearance between coils and magnets\footnote{The requirements on the permanent can be relaxed significantly if more complicated coils (such as saddle coils or shaped modular coils) are allowed, but the basic optimizition proceeds in the same way, seeking to minimize Eq.~(\ref{Mmax}).}. 

\begin{figure}[htp]
   \centerline{\includegraphics[scale=0.5]{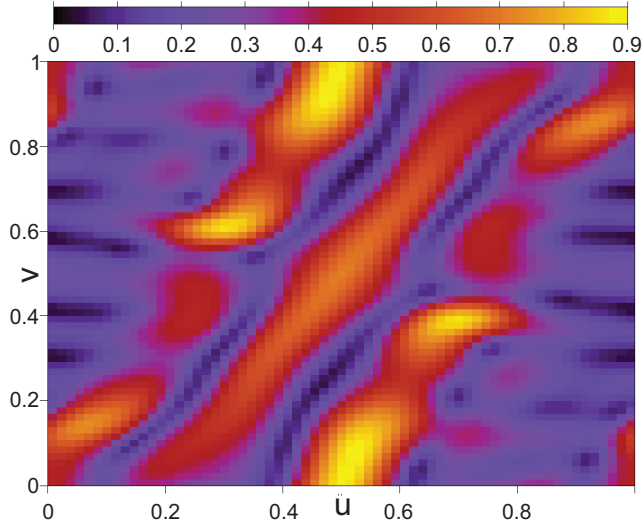}}
	\caption{\em The quantity $\mu_0 |{\bf K}|$ in Tesla, i.e., the magnetization surface current density (\ref{skin current}) multiplied by $\mu_0$, on a surface situated 10 cm from the plasma boundary shown in Figs.~1,3 and 4. The device has been scaled to a field strength on the magnetic axis $B_0 = 1$ T and an average major radius $R=1.4$ m.}
\end{figure}

The fidelity of the magnetic field is very good, as indicated in Fig.~3, which compares flux surfaces with those in the original design. The magnets are essential: the field from the coils alone does not trace out magnetic surfaces. The quality of the quasi-axisymmetry is such that the largest non-axisymmetric harmonics (in Boozer coordinates) of the field strength is only about 1\%. As a result, neoclassical calculations indicate that the effective ripple (the standard measure for transport in the so-called $1/\nu$-regime \cite{Beidler}) remains below 1\% throughout the plasma. Moreover, magnetohydrodynamic Mercier stability is at least as good as the original design in most of the plasma volume. 

\begin{figure}[htp]
   {\includegraphics[scale=0.5]{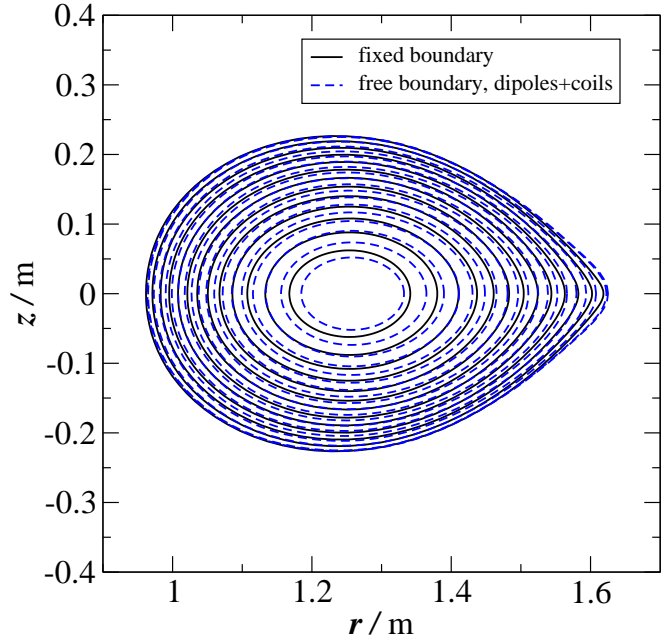}}
	\caption{\em Flux surfaces produced by the coils and magnets shown in Figs.~1-2, and those in the target for the original design, calculated as a fixed-boundary VMEC equilibrium.}
\end{figure}

A concrete arrangement of the permanent magnets follows upon specifying the function $f(r,u,v)$. An illustrative example is shown in Fig.~4, where we have chosen the coordinate $r$ to denote the distance from the surface $\p \Omega_0$ and $f(r,u,v) = 1 - r/d$, with $d = 18$ cm. The figure shows the magnetization vector $\bf M$ on the surface $r = d/2$ as arrows centered on this surface. In certain toroidal and poloidal positions, the required magnetization strength is very small, thus enabling the installation of ports without significantly disturbing the magnetic field.

\begin{figure}[htp]
   \rotatebox{90}{\includegraphics[scale=0.4]{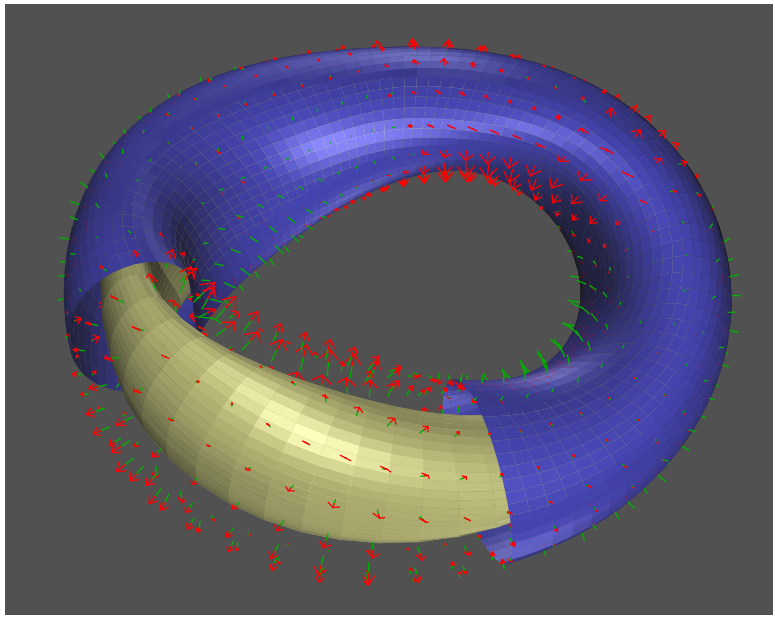}}
	\caption{\em Magnetization vector on a surface in the middle of the magnetization region. This surface is shown in blue and the magnetization vector is represented in the form of arrows centered on this surface. The front half of each arrow and is colored red and the trailing half is green.}
\end{figure}

Further optimization is both possible and desirable. The procedure described above gives a magnetization field $\bf M$ creating the desired magnetic field on the plasma surface (as well as possible), but there are other possible choices of $\bf M$ producing the same magnetic field with fewer or less powerful magnets. Indeed, any field given by Eq.~(\ref{BS}) remains the same if $\bf M$ is replaced by $\tilde {\bf M} = {\bf M} + \nabla \chi$, where $\chi$ is an arbitrary function that vanishes on the boundary $\p \Omega$. This gauge freedom to modify $\bf M$ without changing the magnetic field can be used to minimize the total squared magnetization,
	\bn W = \int_\Omega \tilde M^2 d{\bf r}
	= \int_\Omega \left[ M^2 + (\nabla \chi)^2 - 2 \chi \nabla \cdot {\bf M} \right] d{\bf r}.
	\label{W}
	\en
The function $\chi$ that minimizes $W$ satisfies the Euler-Lagrange equation 
	\bn \nabla^2 \chi = - \nabla \cdot {\bf M} 
	\label{Poisson}
	\en
and thus implies $ \nabla \cdot \tilde {\bf M} = 0$ in $\Omega$. The magnetization field with the smallest value of $W$ is thus divergence-free. Thus, if a small value of $W$ is desired, the magnetization field found from Eq.~(\ref{Merkel M}) can be refined by adding a field $\nabla \chi$ obtained by solving Poisson's equation (\ref{Poisson}) with $\chi = 0$ on the boundary. 

An important technical limit on permanent magnets is given by the maximum value of $\mu_0 M$, which should not exceed about 1.4 T for Nd magnets. To stay below this limit, one could minimize the maximum value of $M$ within $\Omega$ subject to the constraint that Eq.~(\ref{BS}) should be equal to the desired magnetic field on the plasma boundary. It is useful to note that the resulting  magnetization strength cannot have any isolated maxima, for any such maximum can removed by a gauge transformation. To see this, suppose that $M$ has a local maximum at ${\bf r} = {\bf r}_{\rm max}$ and consider the field
	$$ \tilde {\bf M} = {\bf M} - \epsilon \nabla S, 
	$$
where $\epsilon$ is a small positive number and $S({\bf r})$ any differentiable function with bounded derivatives such that
	$$ {\bf M} \cdot \nabla S > 0 $$
in a neighborhood of ${\bf r}_{\rm max}$ and $S({\bf r})=0$ on the boundary $\p \Omega$. As before, $\tilde {\bf M}$ produces the same magnetic field as ${\bf M}$, but 
	$$ \tilde M^2 = M^2 - 2 \epsilon {\bf M} \cdot \nabla S + \epsilon^2 |\nabla S |^2 
	$$
is smaller than $M^2$ in a neighborhood of ${\bf r}_{\rm max}$ if $\epsilon$ is sufficiently small. This result suggests (but does not quite prove) that it should be sufficient to look for magnetization fields with constant amplitude everywhere, which is useful since it reduces the number of free functions in the optimization from three to two and makes optimal use of the available magnetization. (The thickness of the magnetization region can then be reduced correspondingly.) Simple estimates suggest that this kind of optimization should be able to increase the achievable field strength by about a factor of two as compared with Eq.~(\ref{Merkel M}) \footnote{For instance, one may consider the problem of finding the largest field in a point produced by magnets located at distances $r \in [r_1,r_2]$ from this point. With Merkel's method, the result can never exceed $B = \pi \mu_0 M/4$ (and is substantially smaller if $r_1 \simeq r_2)$, whereas an optimally chosen magnetization field with constant amplitude gives $B = 1.38 \mu_0 M \ln(r_2/r_1)$.}. (As in the case of ``one-sided"' magnets, the most efficient use of the magnetization is made if the flux is directed towards one side of the magnetization region \cite{Mallinson}.) Moreover, one finds that, using this type of optimization, the magnetization region can practically be reduced to zero in large regions on the outboard side of the torus, thus providing plenty of room for ports \cite{Landreman}.

Even if the magnitude of the vector $\bf M$ is made constant throughout the magnetization region, its direction will still vary. Any discretization of the magnetic material will thus cause field errors \cite{Choi,Jin}, but these can be made arbitrarily small by making the individual magnets small. The resulting magnetic ripple is exponentially small if the discretization length scale is smaller than the distance to the plasma.

In conclusion, we have shown that it is possible, at least in principle, to use permanent magnets to shape a stellarator plasma. Such magnets cannot produce toroidal magnetic flux, but they can create poloidal flux and rotational transform of the magnetic field, and thus help to simplify stellarator design. In contrast to coils, they can easily be arranged in complicated patterns and do not require power supplies or cooling. They do of course suffer from other disadvantages, such as limitations in field strength, non-tunability, and the possibility of demagnetization. A high-performance stellarator needs a magnetic field much larger than that which present-day permanent magnets can produce, but this does not prevent permanent magnets with $\mu_0 M \le 1.4 $ T to be useful, since such magnets only need to produce a fraction of the full field. How large this fraction is, depends on how much coil complexity can be tolerated. The magnets retain their magnetization in fields up to 5-7 T, and further advances in magnet technology is likely to lead to enhanced performance in the future. Moreover, permanent magnets may bring practical advantages in addition to coil simplification. It would be useful to retain enough flexilibilty in the positioning of the magnets that they could be rotated and repositioned in order to adjust for field imperfections or to create a variety of different magnetic configurations in a single device. In a stellarator with superconducting coils, it could prove possible to reduce the number of coils sufficiently that each coil is situated in its own cryostat, which would tremendously facilitate access to the plasma vacuum vessel. 

The first author would like to acknowledge the splendid hospitality of PPPL, where this work was initiated. This work was supported by a grant from the Simons Foundation (560651, PH).


\end{document}